\documentclass{webofc}
\usepackage[varg]{txfonts}   
\begin{document}
%
\title{Equation of State and Energy Loss of Hot and Dense Quark-Gluon matter from Holographic Black Holes}
%
%

\author{\firstname{Joaquin} \lastname{Grefa}\inst{1}\fnsep\thanks{\email{vgrefaju@central.uh.edu}} \and
        \firstname{Mauricio} \lastname{Hippert}\inst{2} \and
        \firstname{Jorge} \lastname{Noronha}\inst{2} \and
        \firstname{Jacquelyn} \lastname{Noronha-Hostler}\inst{2} \and
        \firstname{Israel} \lastname{Portillo}\inst{1} \and
        \firstname{Claudia} \lastname{Ratti}\inst{1} \and
        \firstname{Romulo} \lastname{Rougemont}\inst{3}
}

\institute{Physics Department, University of Houston, Houston, Texas 77204, USA 
\and
           Illinois Center for Advanced Studies of the Universe, Department of Physics, University of Illinois at Urbana-Champaign, Urbana, Illinois 61801, USA 
\and
           Instituto de F\'{i}sica, Universidade Federal de Goi\'{a}s, Av. Esperan\c{c}a - Campus Samambaia, CEP 74690-900, Goi\^{a}nia, Goi\'{a}s, Brazil
          }

\abstract{%
  By using gravity/gauge correspondence, we construct a holographic model, constrained to mimic the lattice QCD equation of state at zero density, to investigate the temperature and baryon chemical potential dependence of the equation of state. We also obtained the energy loss of light and heavy partons within the hot and dense plasma represented by the heavy quark drag force, Langevin diffusion coefficients and jet quenching parameter at the critical point and across the first-order transition line predicted by the model.
}
\maketitle
\section{Introduction}
\label{intro}

At vanishing density, strongly interacting matter undergoes a smooth but rapid crossover transition from hadrons at low temperature to a system of deconfined quarks and gluons at high temperature \cite{Aoki:2006we}, a strongly interacting liquid called the quark-gluon plasma (QGP). This crossover is expected to evolve into a line of first order phase transition with a critical end point (CEP) at finite baryon chemical potential $\mu_{B}$. Depending on the location of the CEP in the QCD phase diagram, its effects may be probed in relativist heavy ion collisions by looking at the fluctuations of the baryon charge as a function of the center of mass energy \cite{PhysRevLett.107.052301}. On the theory side, lattice simulations at finite baryon density are limited by the sign problem, and an effective field theory is needed to guide the experimental search for the QCD  critical point.

 Such effective approach to describe hot and dense quark-gluon matter must reproduce the lattice QCD equation of state (EoS) at vanishing chemical potential and exhibit nearly inviscid flow behavior, a feature of the QGP. In fact, Gauge/gravity correspondence \cite{Maldacena:1997re,Gubser:1998bc,Witten:1998qj} has successfully been employed to mimic the physics of the QGP around the crossover and allows calculations both in equilibrium as well as out of equilibrium. In this work, we summarize some of our results from Refs. \cite{Grefa:2021qvt,Grefa:2022sav} where we considered a five-dimensional gravitational theory with a real scalar field $\phi$ (the dilaton field), and a potential of the scalar field $V(\phi)$, responsible for the dynamical breaking of conformal symmetry. Effects due to finite chemical potential are taken into account by adding a Maxwell field $A_{\mu}$ and a coupling function $f(\phi)$. This construction defines a holographic Einstein-Maxwell-dilaton (EMD) model.

\section{The EMD holographic model and the EoS at finite density}
\label{EMD}

The bulk EMD action is given by \cite{DeWolfe:2010he,DeWolfe:2011ts,Critelli:2017oub}, 
\begin{equation}\label{eq:action}
   S= \frac{1}{2\kappa_{5}^{2}}\int_{\mathcal{M}_5} d^{5}x\sqrt{-g}\left[R-\frac{(\partial_\mu \phi)^2}{2}-V(\phi)-\frac{f(\phi)F_{\mu\nu}^{2}}{4}\right],
\end{equation}
where $\kappa_{5}^{2}$ is the 5-dimensional gravitational constant, $g_{\mu\nu}$ is the metric tensor, and $R$ is the Ricci scalar. We are interested here in charged isotropic and translationally invariant black hole backgrounds for the EMD fields. The two free functions in the holographic model, $V(\phi)$ and $f(\phi)$, can be dynamically fixed by matching the holographic EoS and the second order baryon susceptibility $\chi_{2}^{B}$ to the corresponding lattice QCD results with 2 + 1 flavours and physical quark masses at $\mu_{B}=0$ from Refs. \cite{Borsanyi:2013bia,Bellwied:2015lba}.




\begin{figure}
\centering
   \includegraphics[width=0.45\textwidth]{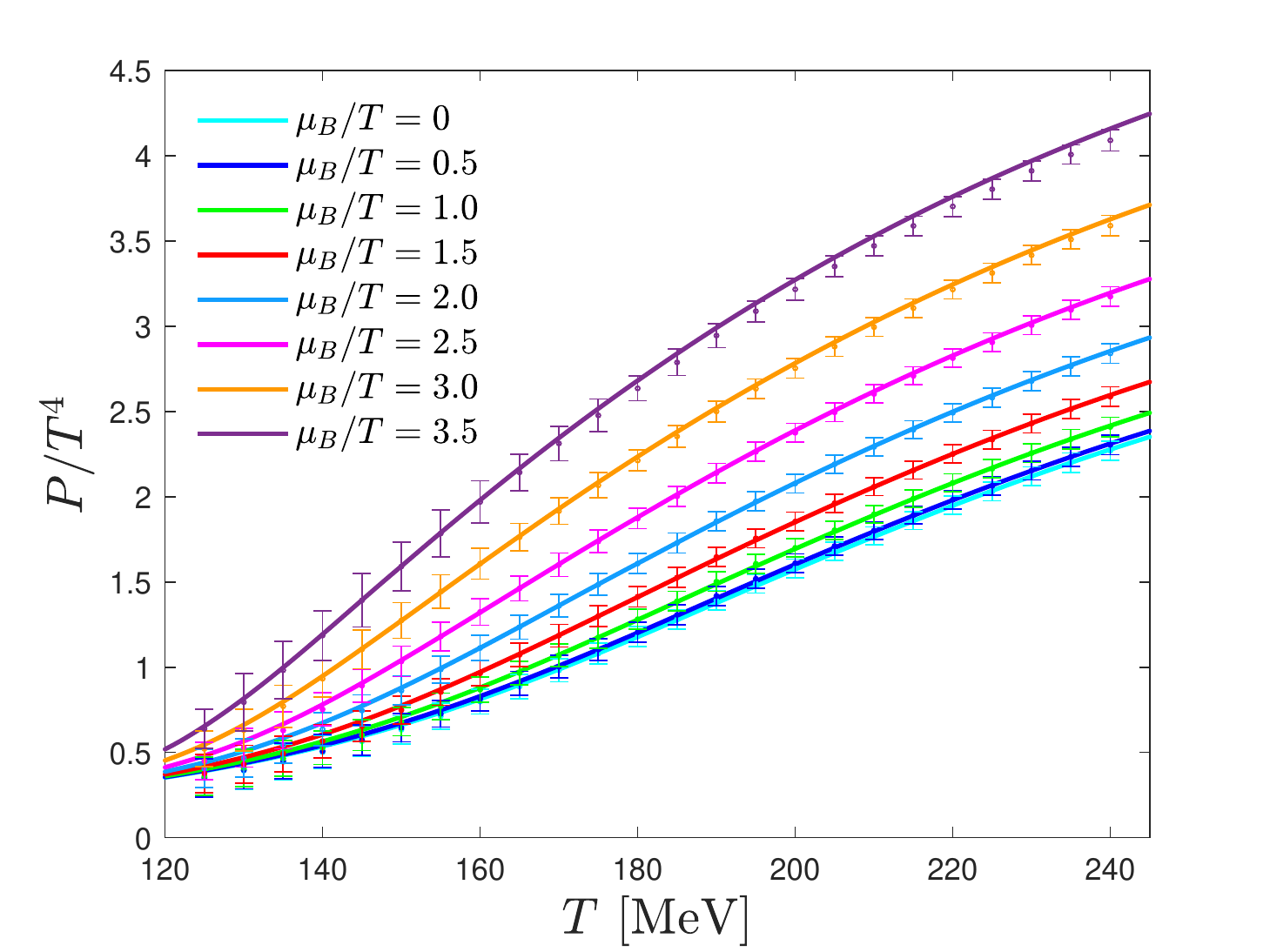}
   \includegraphics[width=0.45\textwidth]{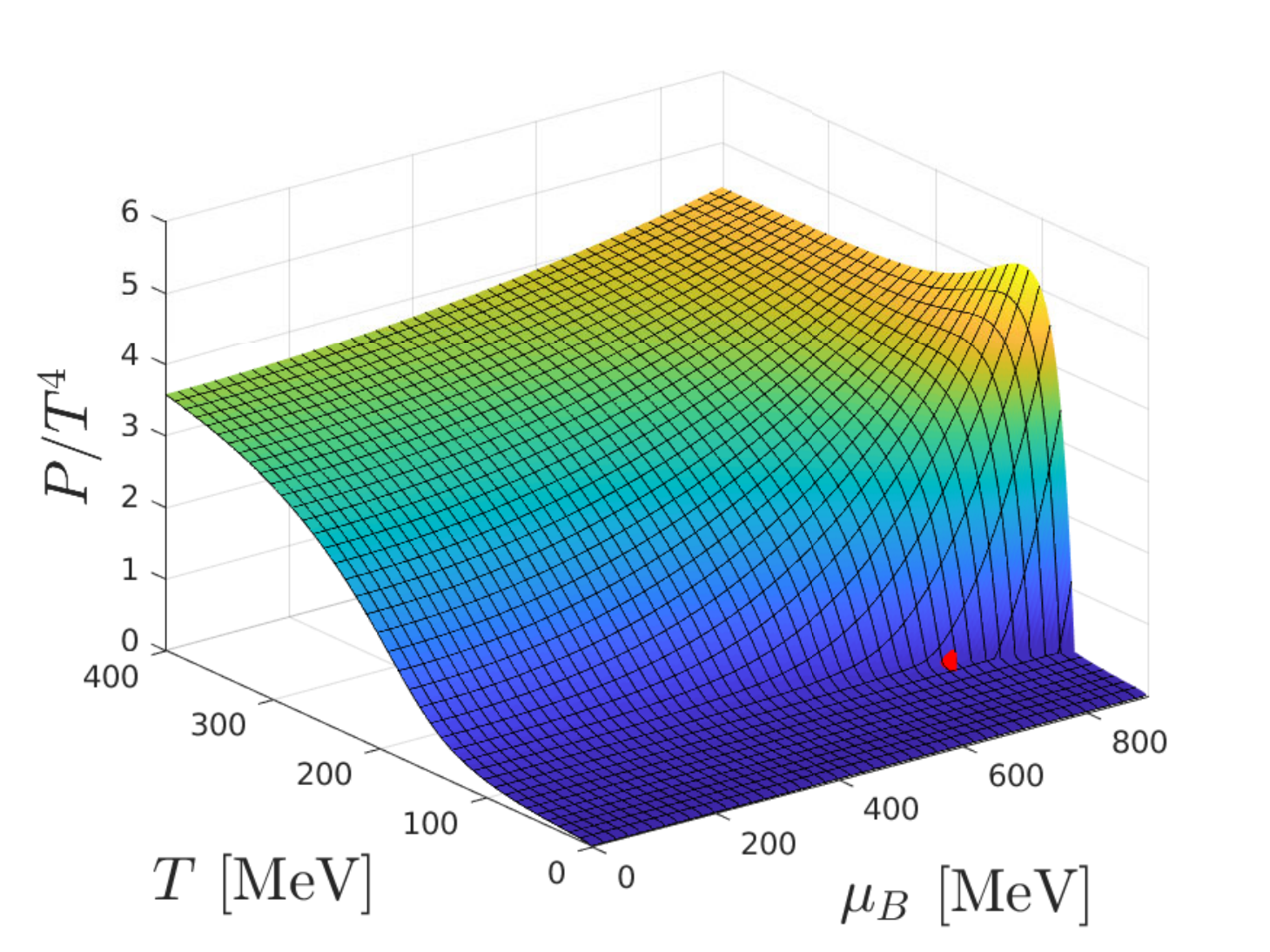}  
\caption{Left: normalized pressure as a function of the temperature for different values of $\mu_{B}/T$ and its comparison with the most recent lattice results. Right, the pressure in the phase diagram with the location of the CEP (red point).}
\vspace{-0.7cm}
\label{fig:Press}       
\end{figure}

The equations of motion obtained from the action (Eq. \ref{eq:action}) are numerically solved given a pair of two initials conditions $(\phi_{0},\Phi_{1})$ where $\phi_{0}$ is the value of the dilaton field at the horizon, and $\Phi_{1}$ is the electric field in the holographic direction evaluated at the horizon. Each black hole solution corresponds to a state in the QCD-like theory with a value for the entropy density ($s$), and baryon density ($\rho_{B}$) over the temperature and baryon chemical potential coordinates computed accordingly to the holographic dictionary \cite{Critelli:2017oub,Grefa:2021qvt}.


The comparison of our results for the holographic pressure with the most recent lattice QCD data up to the ratio of $\mu_{B}/T=3.5$ \cite{Borsanyi:2021sxv} is shown in Fig. \ref{fig:Press}, where we also present this observable over a broad region in the phase diagram with the position of the CEP. The baryon susceptibilities, defined as $\chi_{n}(T,\mu_{B})=\partial^{n}(P/T^{4})/\partial(\mu_{B}/T)^{n}$, diverge in the vicinity of the critical point. In particular, the second order baryon susceptibility $\chi_{2}$ develops a peak at larger chemical potential which evolves into a divergence at the CEP, located at $T^{CEP}=89$ MeV and $\mu_{B}^{CEP}=724$ MeV in this model.

\section{Energy Loss}
\label{Energy_loss}
\begin{figure}[h]
\centering
   \includegraphics[width=0.45\textwidth]{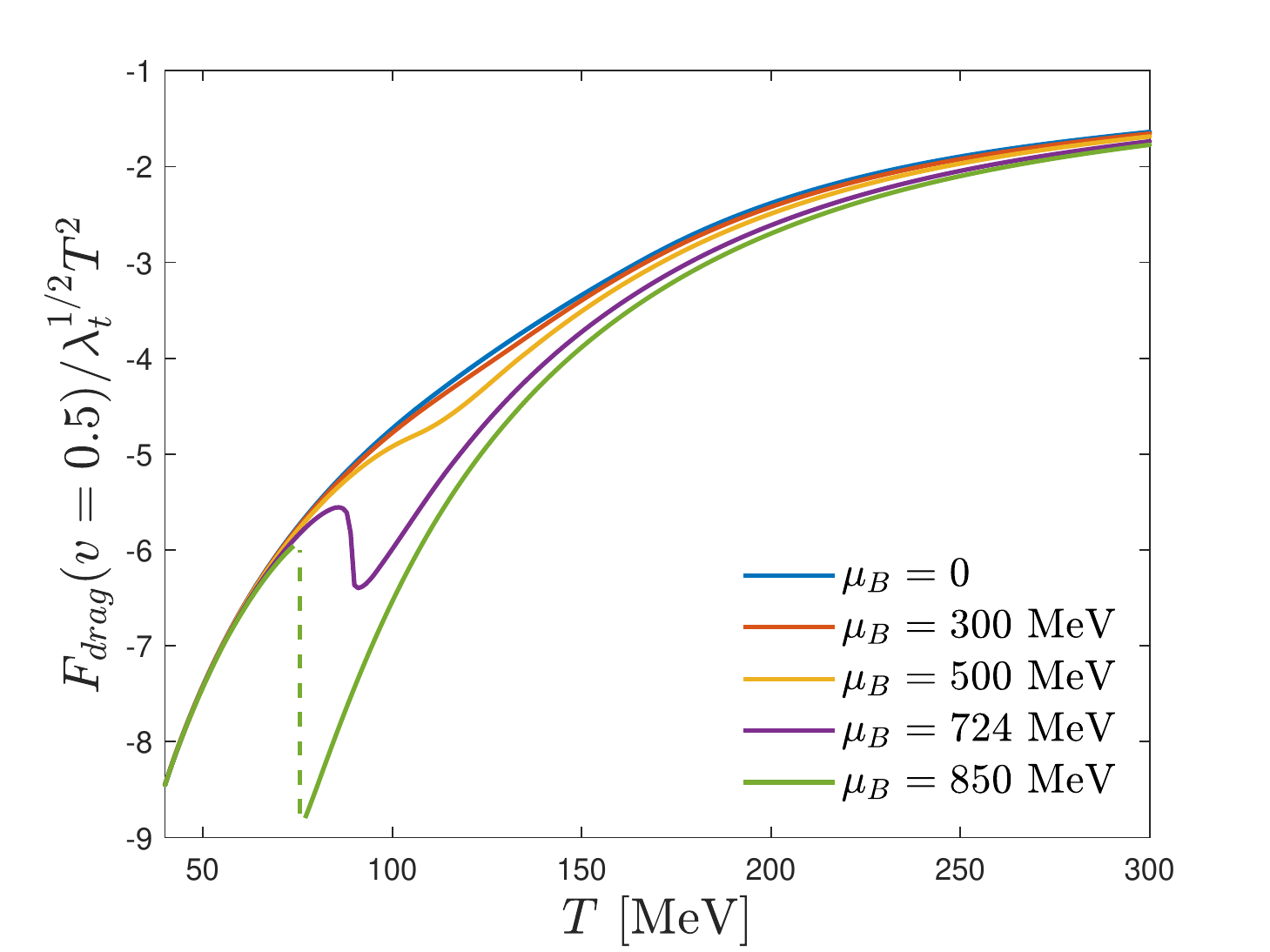}
   \includegraphics[width=0.45\textwidth]{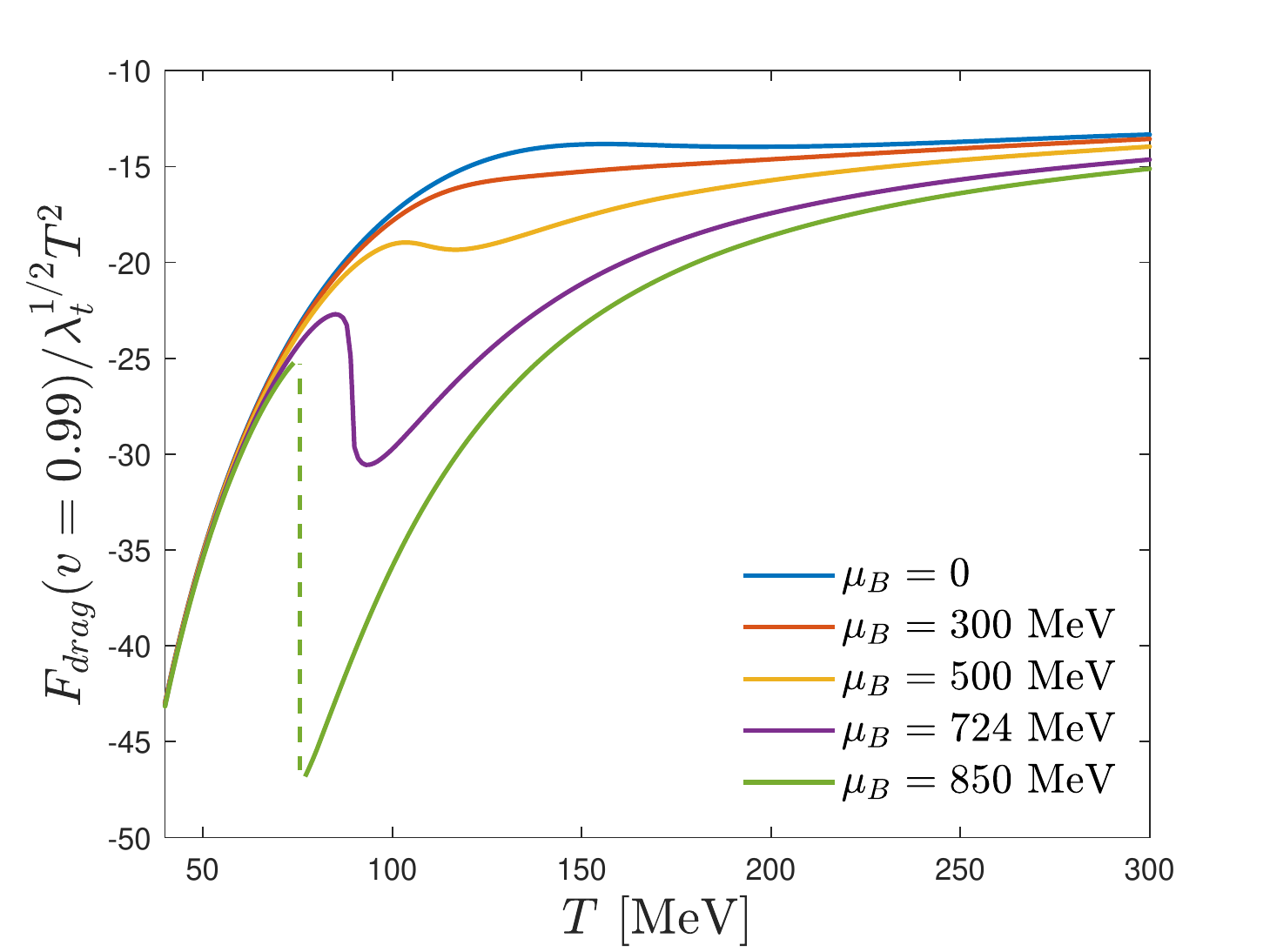}  
\caption{Heavy quark drag force at $v=0.5$ (left) and $v=0.99$ (right) as a function of the temperature for several values of $\mu_{B}$}
\vspace{-0.5cm}
\label{fig:F_drag}       
\end{figure}

As a heavy quark moves through a hot and baryon dense medium, it loses energy and momentum through the drag force $F_{\text{drag}}=dp_{x}/dt$ which can be computed from the holographic trailing string approach \cite{Grefa:2022sav}. The results for the heavy quark drag force for two different values of the quark velocity are shown in Fig. \ref{fig:F_drag}. One can infer that a very heavy quark (i.e. the bottom), which might not achieve a very high velocity within the plasma (the case for $v=0.5$), is less sensitive to the in-medium effects in comparison with a less massive quark (i.e. the charm), which could attain higher velocities within the fluid (the case for $v=0.99$). Additionally, one can observe that the magnitude of the energy loss associated with the heavy quark force increases by lowering the temperature or/and increasing $\mu_{B}$.

\begin{figure}
\centering
   \includegraphics[width=0.45\textwidth]{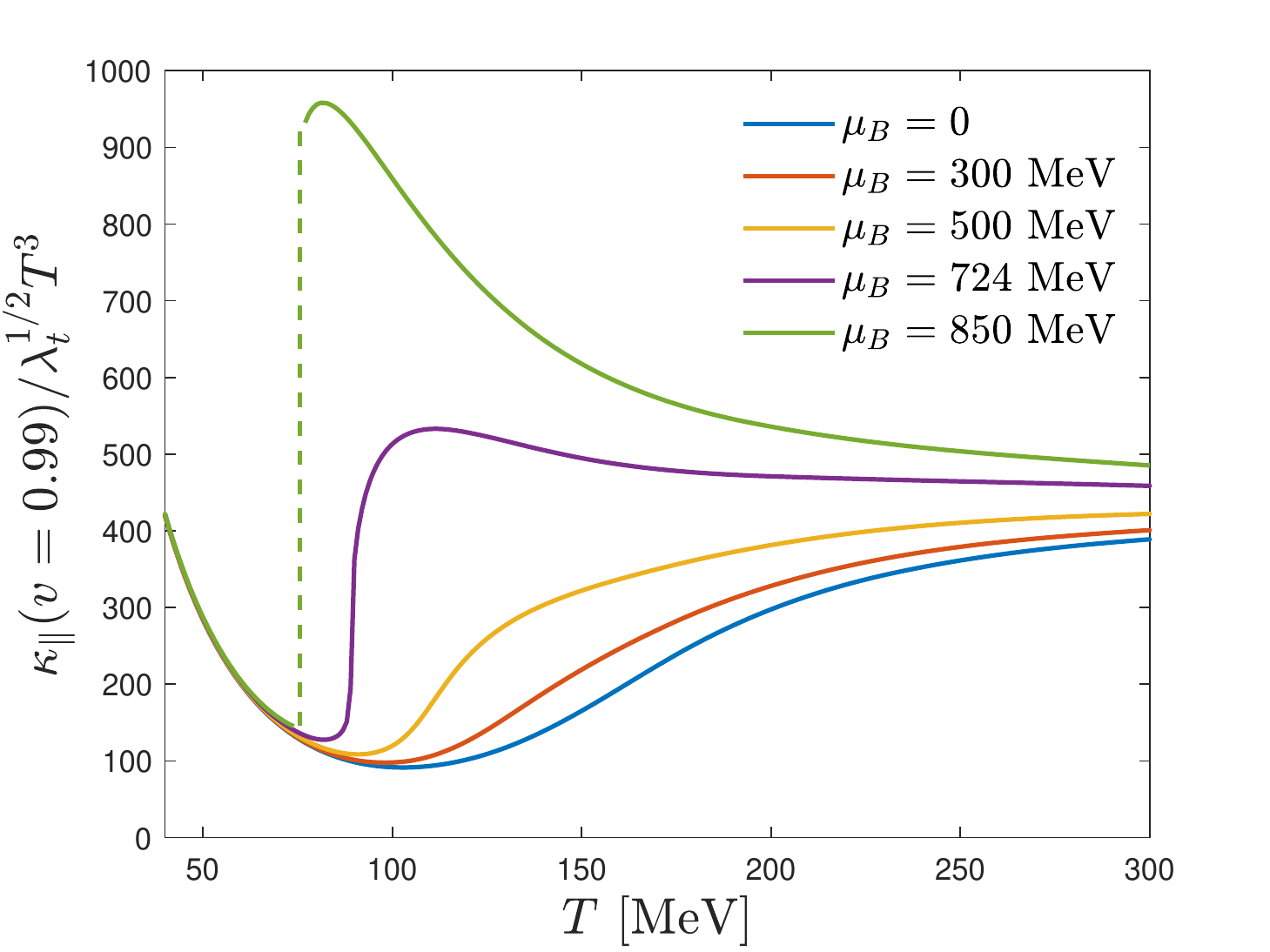}
   \includegraphics[width=0.45\textwidth]{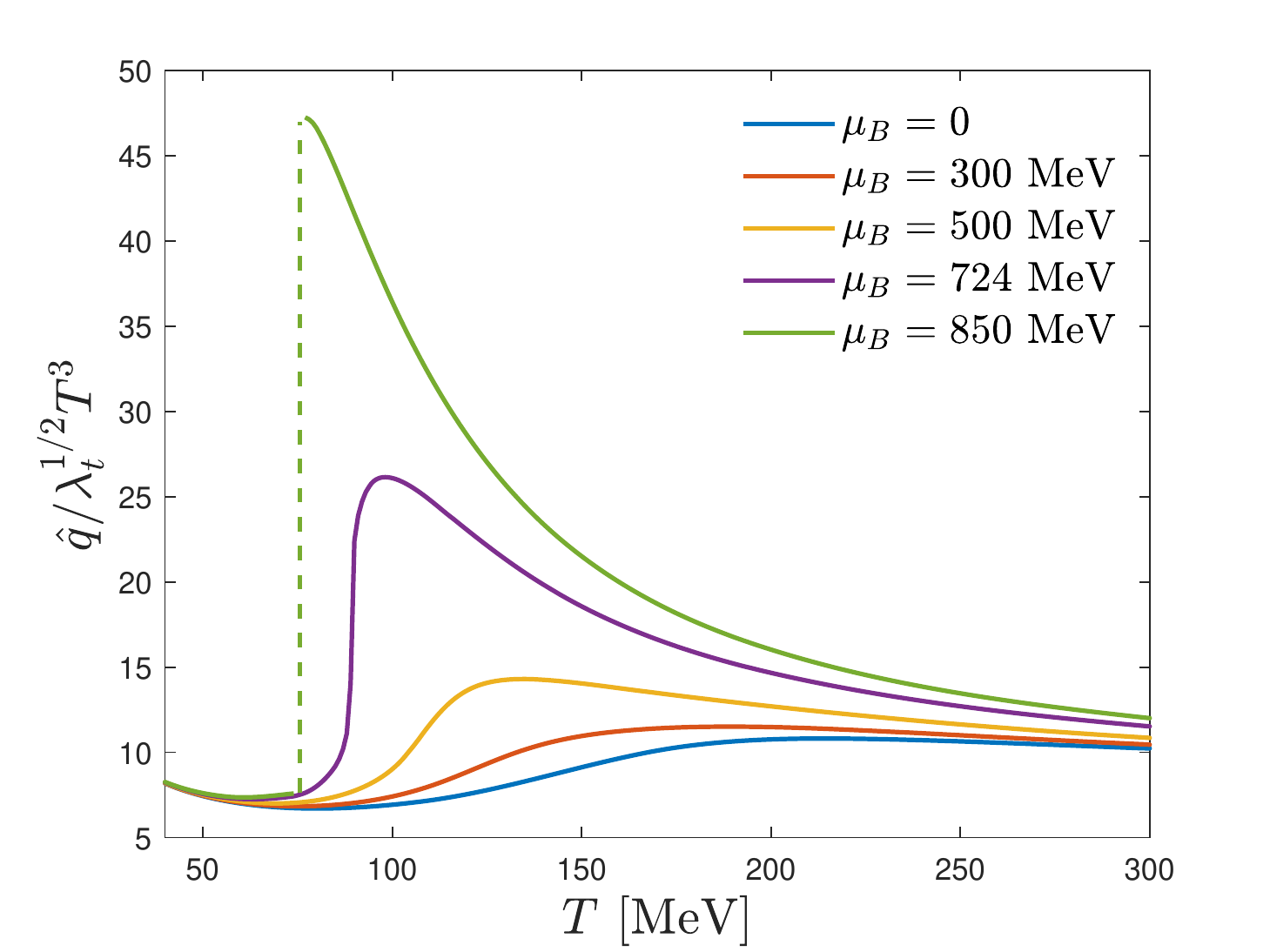}  
\caption{Parallel Langevin diffusion at coefficient $v=0.99$ (left panel) and normalized jet quenching parameter $\hat{q}$ (right panel) as a function of the temperature for several values of $\mu_{B}$.}
\vspace{-0.5cm}
\label{fig:qhat}       
\end{figure}

Langevin diffusion processes can be treated holographically. In particular, the Langevin diffusion coefficients, that describe the thermal fluctuations of a heavy quark trajectory under Brownian motion, can be computed from our EMD setup. The results for the parallel Langevin diffusion coefficient at $v=0.99$ is shown in the left panel of Fig. \ref{fig:qhat}. Analogously to the case of the heavy quark drag force, the Langevin diffusion coefficient is also enhanced with increasing baryon chemical potential. 


Another transport coefficient that can be computed from the holographic EMD model is the jet quenching parameter ($\hat{q}$), defined as the rate for transverse momentum broadening. The jet quenching parameter characterizes the energy loss from collisional and radiative processes of high energy partons produced by the interaction with the hot and dense medium they travel through. The holographic results for the jet quenching parameter are shown in the right panel of Fig. \ref{fig:qhat}. We observe that $\hat{q}/T^{3}$ displays a peak around the crossover that becomes sharper and increases in magnitude as the baryon chemical potential increases, which indicates more jet suppression and parton energy loss in the baryon rich regime. 

In overall, the transport coefficients presented in this contribution exhibit a non trivial dependence with respect to the temperature and chemical potential, and they remain finite at the critical point as expected for holographic / large-$N_c$ approaches (type-B dynamical universality class) \cite{Grefa:2022sav}. While some possess a peak and others a local minimum, it is the inflection point, that gives rise to these extrema, that moves toward the CEP as $\mu_{B}$ is increased, acquiring an infite slope in the critical region. Since these transport coefficients also appear to be sensitive to the transition from the confined to the deconfined phase, its inflection points can be used to define pseudo-transition temperatures over the crossover region. Analogously to the equilibrium variables like the entropy density and baryon density, the transport coefficients display a discontinuity that corresponds to the line of first order phase transition. 

\section{Conclusions}

By using the EMD model from Refs. \cite{Critelli:2017oub,Grefa:2021qvt}, we obtain the equilibrium and dynamical properties for a hot a dense QGP. Analogously to the EoS variables, the transport coefficients presented in this work display a temperature and baryon chemical potential dependence with different inflection points that can be used to describe the crossover transition region. They develop an infinite slope at the CEP that becomes a discontinuity gap at the first order transition line.

\subsection*{Acknowledgements:} 

This material is based upon work supported by the National Science Foundation under grants No. PHY-1654219, PHY-2208724 and PHY-2116686. This work was supported in part by the National Science Foundation (NSF) within the framework of the MUSES collaboration, under grant No. OAC-2103680, the US-DOE Nuclear Science Grant No. DE-SC0020633. J.N. is partially supported by the U.S. Department of Energy, Office of Science, Office for Nuclear Physics under Award No. DE-SC0021301.

%
\bibliography{references.bib}
%
%
%
%

\end{document}